\documentclass[preprint,12pt]{elsarticle}
\usepackage{amssymb}
\usepackage{fullpage}
\usepackage{float}
\restylefloat{table}
\biboptions{sort&compress}

\usepackage{amsmath}
\usepackage{graphicx}
\usepackage[dvipsnames]{xcolor}
\usepackage[colorlinks]{hyperref}
\usepackage{comment}
\usepackage{epstopdf}
\usepackage{subfig}

\def\bR{{\mathbf{R}}}
\def\bx{{\mathbf{x}}}

\journal{arXiv}

\begin{document}

\begin{frontmatter}



\title{SPARC: Simulation Package for Ab-initio Real-space Calculations}


\author[gatech]{Qimen Xu}
\author[gatech]{Abhiraj Sharma}
\author[gatech]{Benjamin Comer}
\author[gatech2]{Hua Huang}
\author[gatech2]{Edmond Chow}
\author[gatech]{Andrew J. Medford}
\author[llnl]{John E. Pask}
\author[gatech]{Phanish Suryanarayana\corref{cor}}

\address[gatech]{College of Engineering, Georgia Institute of Technology, Atlanta, GA 30332, USA}
\address[gatech2]{College of Computing, Georgia Institute of Technology, Atlanta, GA 30332, USA}
\address[llnl]{Physics Division, Lawrence Livermore National Laboratory, Livermore, CA 94550, USA \vspace{-5mm} }
\cortext[cor]{Corresponding Author (\it phanish.suryanarayana@ce.gatech.edu) }

\begin{abstract}
%Ca. 100 words 
We present SPARC: Simulation Package for Ab-initio Real-space Calculations. SPARC can perform Kohn-Sham density functional theory calculations for isolated systems such as molecules as well as extended systems such as crystals and surfaces, in both static and dynamic settings. It is straightforward to install/use and highly competitive with state-of-the-art planewave codes, demonstrating comparable performance on a small number of processors and increasing advantages as the number of processors grows. 
Notably, SPARC brings solution times down to a few seconds for systems with $\mathcal{O}(100-500)$ atoms on large-scale parallel computers, outperforming planewave counterparts by an order of magnitude and more.
\end{abstract}


\end{frontmatter}

\section*{\vspace{-1mm} Code Metadata}
\label{sec:metadata}

\begin{table*}[h!]
\centering
\vspace{-4mm}
\begin{tabular}{|l|p{6.5cm}|p{6.5cm}|}
\hline
C1 & Current code version & v1.0.0 \\
\hline
C2 & Permanent link to code/repository used for this code version & \url{https://github.com/SPARC-X/SPARC}\\
\hline
C3 & Code Ocean compute capsule & \\
\hline
C4 & Legal Code License   & GNU General Public License v3.0 \\
\hline
C5 & Code versioning system used & git \\
\hline
C6 & Software code languages, tools, and services used & C, MPI, BLAS, LAPACK, ScaLAPACK (optional), MKL (optional) \\
\hline
C7 & Compilation requirements, operating environments \& dependencies & OS: Unix, Linux, or MacOS \\
\hline
C8 & If available Link to developer documentation/manual & \url{https://github.com/SPARC-X/SPARC/tree/master/doc} \\
\hline
C9 & Support email for questions & phanish.s@gmail.com \\
\hline
\end{tabular}
\caption{Code metadata}
\label{metadata} 
\end{table*}

% \linenumbers


% The permanent link to code/repository or the zip archive should include the following requirements: 

% README.txt and LICENSE.txt.

% Source code in a src/ directory, not the root of the repository.

% Tag corresponding with the version of the software that is reviewed.

% Documentation in the repository in a docs/ directory, and/or READMEs, as appropriate.

\newpage

\section{Motivation and significance}
\label{sec:motivation}
Over the course of the past few decades, quantum mechanical calculations based on Kohn-Sham density functional theory (DFT) \cite{Kohn1965,hohenberg1964inhomogeneous} have become a cornerstone of materials research by virtue of the predictive power and fundamental insights they provide. The widespread use of the methodology can be attributed to its generality, simplicity, and high accuracy-to-cost ratio relative to other such ab initio approaches \cite{burke2012dft,becke2014dft}.  However, while less expensive than wavefunction based methods, the solution of the Kohn-Sham equations remains a formidable task. In particular, the computational cost scales cubically with the number of atoms, severely limiting the range of physical systems accessible to such first principles investigation. These limitations become even more acute in quantum molecular dynamics (QMD) simulations, wherein the equations for the electronic ground state may be solved tens or hundreds of thousands of times to reach time scales relevant to phenomena of interest \cite{burke2012dft}.

The planewave pseudopotential method \cite{Martin2004} has been among the most widely used techniques for the solution of the Kohn-Sham problem \cite{VASP,CASTEP,ABINIT,Espresso,CPMD,DFT++,gygi2008architecture,valiev2010nwchem}. The underlying Fourier basis is complete, orthonormal, diagonalizes the Laplacian, and provides spectral convergence for smooth problems. As a result, the planewave method is accurate, simple to use since it relies on a single convergence parameter, and highly efficient on moderate computational resources with the use of well optimized Fast Fourier Transforms (FFTs) and efficient  preconditioning schemes. However, the Fourier basis restricts the method to periodic boundary conditions, whereby finite systems such as clusters and molecules, as well as semi-infinite systems such as surfaces and nanowires, require the introduction of artificial periodicity with large vacuum regions. This limitation also necessitates the introduction of an unphysical neutralizing background density when treating charged systems in order to avoid Coulomb divergences. Moreover, the global nature of the Fourier basis hampers scalability on parallel computing platforms and complicates the development of linear-scaling methods \cite{Goedecker,Bowler2012,aarons2016perspective}, limiting the system sizes and time scales accessible.

The limitations of the planewave method have motivated the development of a number of alternative solution strategies employing systematically improvable, localized representations \cite{becke1989basis,chelikowsky1994finite,genovese2008daubechies,seitsonen1995real,white1989finite,iwata2010massively,tsuchida1995electronic,xu2018discrete,Phanish2011,Phanish2010,ONETEP,CONQUEST,MOTAMARRI2020106853,OCTOPUS,briggs1996real,fattebert1999finite,shimojo2001linear,ghosh2017sparc2,arias1999wav,pask2005femeth,lin2012adaptive}. Among these, perhaps the most mature and widely used to date are the finite-difference methods \cite{beck2000rsmeth,saad2010esmeth}, wherein computational locality is maximized by discretizing all quantities of interest on a uniform real-space grid. As a result, convergence is controlled by a single parameter and both periodic and Dirichlet boundary conditions are naturally accommodated, thus enabling the efficient and accurate treatment of finite, semi-infinite, bulk, and charged systems alike.  Moreover,  real-space methods are amenable to the development of linear scaling methods, and large-scale parallel computational resources can be efficiently leveraged by virtue of the method's simplicity, locality, and freedom from communication-intensive transforms such as FFTs \cite{shimojo2001linear,iwata2010massively,hasegawa2011first,osei2014accurate,suryanarayana2018sqdft}. With these and other advances, real-space methods have been applied to systems containing thousands of atoms, and have demonstrated substantially reduced solution times compared to established planewave codes in applications to both finite \cite{ghosh2017sparc1} and extended \cite{ghosh2017sparc2} systems. 

However, despite the significant advantages afforded by real-space methods, the planewave method has remained the method of choice in practice for the better part of the past two decades. This is largely due to the ease of use, extensive feature sets, established accuracy/robustness, and straightforward installation of the associated codes, having been in development and production for a longer period of time. Perhaps most importantly, however, planewave codes have typically yielded shorter times to solution using moderate computational resources, as most widely available to researchers in practice \cite{ghosh2017sparc1,ghosh2017sparc2}. Moreover, even with access to larger-scale machines, real-space codes have not always yielded shorter times to solution, further hindering wider adoption in practice. 

In this work, we present an open-source software package for the accurate, efficient, and scalable solution of the Kohn-Sham equations, referred to as SPARC. The package is straightforward to install/use and highly competitive with state-of-the-art planewave codes, demonstrating comparable performance on a small number of processors and order-of-magnitude advantages as the number of processors increases.

%%%%%%%%%%%%%%%%%%%%%%%%%%%%%%%%%%%%%%%%%%%%%%%%%%%%%%%%%%%%%%%%%%%%%%%%%%%%%%%%%%%%%%%%%%%%%
%%%%%%%%%%%%%%%%%%%%%%%%%%%%%%%%%%%%%%%%%%%%%%%%%%%%%%%%%%%%%%%%%%%%%%%%%%%%%%%%%%%%%%%%%%%%%
%%%%%%%%%%%%%%%%%%%%%%%%%%%%%%%%%%%%%%%%%%%%%%%%%%%%%%%%%%%%%%%%%%%%%%%%%%%%%%%%%%%%%%%%%%%%%

\section{Software description} 
\label{sec:description}
%Describe the software in as much as is necessary to establish a vocabulary needed to explain its impact. 

The central focus of SPARC is the  accurate and efficient solution of the finite-temperature Kohn-Sham equations for the electronic ground state \cite{Kohn1965,hohenberg1964inhomogeneous,Mermin1965}:
\begin{equation} \label{Eqn:KSeq}
\left( \mathcal{H}^{\sigma} \equiv - \, \frac{1}{2} \nabla^2 + V_{\rm eff}^{\sigma} \left[ \rho^{\alpha}, \rho^{\beta}; \bR \right] \right)  \psi_n^{\sigma}  =  \lambda_n^{\sigma} \psi_n^{\sigma} \,, \quad n=1,2, \ldots, N_s^{\sigma} \,, \quad \sigma \in \{\alpha, \beta\} \,,
\end{equation}
where the superscript $\sigma$ denotes the spin, i.e., spin-up or spin-down, $\mathcal{H}^{\sigma}$ is the Hamiltonian, $\psi_n^\sigma$ are the orthonormal orbitals with energies $\lambda_n^{\sigma}$, $V_{\rm eff}^{\sigma}$ is the effective potential, $N_s^{\sigma}$ is the number of states, and $\bR$ denotes the set of atomic positions. In addition, $\rho^{\sigma}$ represents the spin-resolved electron density:
\begin{equation} \label{Eqn:SpinElectronDensity}
\rho^{\sigma}(\bx) = \sum_{n=1}^{N_s^{\sigma}}  g_n^{\sigma} {|\psi_n^{\sigma}(\bx)|}^2 \,, \quad \sigma \in \{\alpha,\beta\} \,, \quad \bx \in \mathbb{R}^3 \,,
\end{equation}
where $g_n^{\sigma}$ are the orbital occupations, typically given by the Fermi-Dirac distribution. In implementations of the above equations, once a suitable fundamental domain/unit cell has been identified, zero Dirichlet or Bloch-periodic boundary conditions are prescribed on the orbitals along the directions in which the system is finite or extended, respectively.

%%%%%%%%%%%%%%%%%%%%%%%%%%%%%%%%%%%%%%%%%%%%%%%%%%%%%%%%%%%%%%%%%%%%%%%%%%%%%%%%%%%%%%%%%%%%
%%%%%%%%%%%%%%%%%%%%%%%%%%%%%%%%%%%%%%%%%%%%%%%%%%%%%%%%%%%%%%%%%%%%%%%%%%%%%%%%%%%%%%%%%%%%

\subsection{Software Architecture }
\label{subsection:architecture}

\begin{figure}[htbp]
\centering
\includegraphics[keepaspectratio=true,width=0.72\textwidth]{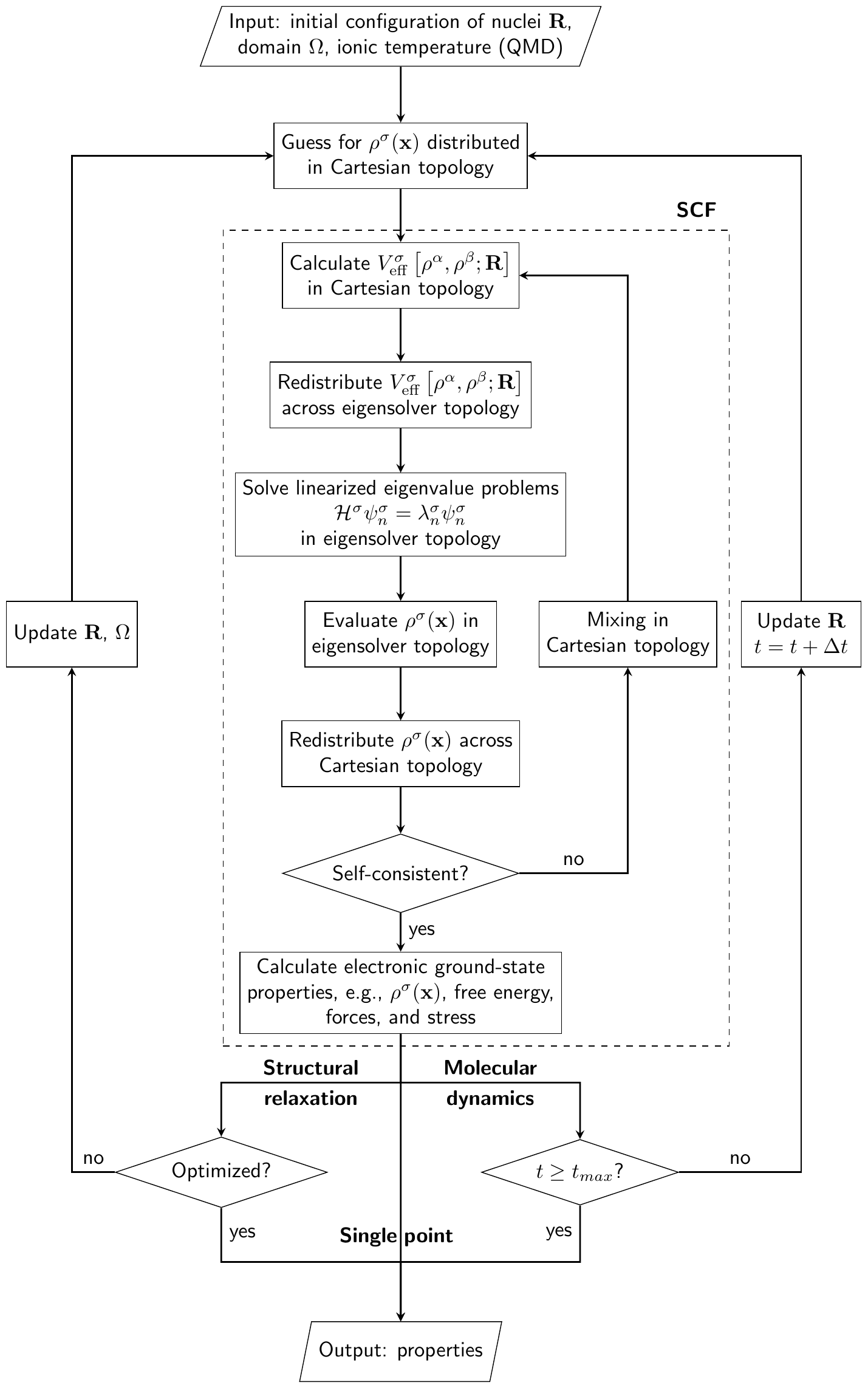}
\caption{Overview of the SPARC framework for performing Kohn-Sham  DFT calculations. The Cartesian topology is formed by embedding a three-dimensional processor grid  into the \texttt{MPI\_COMM\_WORLD} communicator. The eigensolver topology is a collection of smaller Cartesian topologies, created by first splitting the \texttt{MPI\_COMM\_WORLD} communicator into multiple spin groups, then splitting each spin group into multiple Brillouin zone integration groups, then splitting each Brillouin zone integration group into multiple band groups, and finally, embedding each band group with a Cartesian topology.}
\label{Fig:OverallFlowChart}
\end{figure}

SPARC employs the pseudopotential approximation \cite{Martin2004} to facilitate the efficient solution of the Kohn-Sham equations for the whole of the periodic table of elements.  In addition, it  employs a local real-space formulation of the electrostatics  \cite{Suryanarayana2014524,ghosh2014higher}, wherein the electrostatic potential --- component of $V_{\rm eff}^{\sigma}$ that is the sum of ionic and Hartree contributions --- is given by the solution of a Poisson problem, with Dirichlet or periodic boundary conditions prescribed along directions in which the system is finite or extended, respectively.  In this framework, SPARC performs a uniform real-space discretization of the equations, using a high-order centered finite-difference approximation for differential operators and the trapezoidal rule for integral operators. The actual code is written in the \texttt{C} language and achieves parallelism through the message passing interface 
(\texttt{MPI}) \cite{gropp1999using}. 
An overview of the SPARC framework for performing Kohn-Sham DFT calculations is shown in Fig.~\ref{Fig:OverallFlowChart}. 

SPARC can perform 
single-point calculations, structural relaxations (atom and/or cell), and QMD simulations. 
For single-point calculations, the electronic ground state is determined for fixed ionic positions and cell dimensions, whereas for structural relaxations, positions and/or cell dimensions are varied to minimize the Kohn-Sham energy using the Hellmann-Feynman atomic forces \cite{ghosh2017sparc1,ghosh2017sparc2} and/or stress tensor \cite{sharma2018calculation}. For QMD, the ionic positions, velocities, and accelerations are evolved by integrating the equations of motion, with or without a thermostat, using the atomic forces. In all cases, the calculations can be either spin-polarized or unpolarized, with various choices of local and semilocal exchange-correlation functionals. 

SPARC requires two input files for every calculation: (i) a \texttt{.inpt} file containing the options and parameters to be used in the calculation, including the choice of exchange-correlation functional, flag for spin-polarization, type of static/dynamic calculation, ionic temperature in the case of QMD,  cell dimensions, boundary condition in each direction, and finite-difference grid specification;  and (ii) a \texttt{.ion} file containing information on the atomic configuration, including atom types, positions, and paths to corresponding pseudopotential files. Note that, in order to enable detailed control of the simulation, a large number of parameters and options can be specified, as described in the accompanying user guide. However, by virtue of carefully chosen defaults, relatively few parameters typically need be specified in practice. Note also that, since all files are simple, human-readable text, series of simulations are readily scripted. 
A Python package containing helper functions for generating input files and submitting simulations is also available.

The Kohn-Sham problem for the electronic ground state needs to be  solved for every  configuration of atoms encountered during the DFT simulation. In SPARC, this is achieved using the self-consistent field (SCF) method \cite{Martin2004}, which represents a fixed-point iteration with respect to either the electron density or potential. For the first SCF iteration in the simulation,  a superposition of isolated atom electron densities is used as the initial guess, whereas for every subsequent atomic configuration encountered, extrapolation based on solutions to  previous configurations is employed \cite{alfe1999ab}. The convergence of the SCF iteration is accelerated using the  restarted variant \cite{pratapa2015restarted} of the Periodic Pulay mixing scheme \cite{banerjee2016periodic} with a real-space preconditioner \cite{kumar2019preconditioning}. In the case of spin-polarized calculations, mixing is performed simultaneously on both components, i.e., on a vector of twice the original length containing both spin-up and spin-down density/potential components. 

In each SCF iteration, SPARC performs a partial diagonalization (i.e., eigenvalues and eigenvectors calculated approximately) of the linear eigenproblem using the CheFSI method \cite{zhou2006self,zhou2006parallel}, with multiple Chebyshev filtering steps performed in the first iteration of the simulation \cite{zhou2014chebyshev}. The Hamiltonian-matrix/vector products are performed in matrix-free fashion, using the finite-difference stencil for the Laplacian and the outer product nature of the nonlocal pseudopotential operator. While doing so, zero-Dirichlet or Bloch-periodic boundary conditions are prescribed on the orbitals along directions in which the system is finite or extended, respectively. In calculating the effective potential, the Poisson problem for the electrostatic potential is solved using the AAR method \cite{pratapa2016anderson,suryanarayana2019alternating}, with Laplacian-vector products again performed in matrix-free fashion using the finite-difference stencil. While doing so, Dirichlet or periodic boundary conditions are prescribed on the electrostatic potential along directions in which the system is finite or extended, respectively. In particular, Dirichlet values are determined using a multipole expansion for isolated systems and a dipole correction for surfaces and nanowires \cite{burdick2003parallel,natan2008real}.

In SPARC,  information pertaining to the overall DFT simulation is written to the \texttt{.out} file, including progression of the SCF iteration, electronic ground state energy, maximum  atomic force, maximum  stress, and various timings. Based on the type of  calculation, a \texttt{.static}, \texttt{.geopt}, or \texttt{.aimd} file may also be written. The \texttt{.static} file  contains information about the single-point calculation, including atom positions, electronic ground state energy, forces, and stress tensor. The \texttt{.geopt} file contains information about the structural relaxation, including (i) atom positions, electronic ground state energy, and  forces (atomic relaxation), and (ii) cell information and  stress tensor (cell relaxation). The \texttt{.aimd} file contains information about the QMD simulation, including atom positions, forces, and velocities. To seamlessly continue from a previously stopped simulation, a \texttt{.restart} file is written for structural relaxation and QMD calculations. SPARC provides other outputs if specified as well, e.g., a \texttt{.eigen} file containing eigenvalues and occupations and \texttt{.dens} file containing the charge density. 

%%%%%%%%%%%%%%%%%%%%%%%%%%%%%%%%%%%%%%%%%%%%%%%%%%%%%%%%%%%%%%%%%%%%%%%%%%%%%%%%%%%%%%%%%%%%
%%%%%%%%%%%%%%%%%%%%%%%%%%%%%%%%%%%%%%%%%%%%%%%%%%%%%%%%%%%%%%%%%%%%%%%%%%%%%%%%%%%%%%%%%%%%

\subsection{Software Functionalities}
\label{subsection:functionalities}
The current version of SPARC is capable of performing spin-polarized and unpolarized  ab initio calculations based on Kohn-Sham DFT for isolated systems such as molecules as well as extended systems such as crystals, surfaces, and nanowires, in both static and dynamic settings. Specifically, it can  perform single-point calculations for a given atomic configuration, structural relaxations with respect to atom positions and/or cell dimensions \cite{NLCG, LBFGS, FIRE,press2007numerical}, and NVE/NVT/NVK  QMD simulations \cite{NVE, NVTNH, NVK}.  Available exchange-correlation functionals include various forms of LDA \cite{PhysRevB.23.5048,perdew1992accurate} and  GGA \cite{perdew1996generalized,RPBE,PBEsol}. Types of pseudopotentials employed are ONCV \cite{hamann2013optimized}  and Troullier-Martin \cite{Troullier}, both in \texttt{psp8} format \cite{ABINIT}. Over the course of simulations, in addition to electronic density and free energy, SPARC can calculate atomic forces, pressure, and the stress tensor for extended systems. The outputs from such DFT calculations can be used to calculate a number of properties, including lattice constant, cohesive energy, polarization, elastic moduli, density of states, electronic band structure, pair distribution function, equations of state, shear viscosity, defect energy, surface energy, absorption energy, equilibrium bond lengths, HOMO-LUMO gap, and dipole moment. 

%%%%%%%%%%%%%%%%%%%%%%%%%%%%%%%%%%%%%%%%%%%%%%%%%%%%%%%%%%%%%%%%%%%%%%%%%%%%%%%%%%%%%%%%%%%%
%%%%%%%%%%%%%%%%%%%%%%%%%%%%%%%%%%%%%%%%%%%%%%%%%%%%%%%%%%%%%%%%%%%%%%%%%%%%%%%%%%%%%%%%%%%%
%%%%%%%%%%%%%%%%%%%%%%%%%%%%%%%%%%%%%%%%%%%%%%%%%%%%%%%%%%%%%%%%%%%%%%%%%%%%%%%%%%%%%%%%%%%%

\section{Illustrative Examples}
\label{sec:examples}
%%Provide at least one illustrative example to demonstrate the major functions.

We now demonstrate  the major functionalities of SPARC through examples representative of physical applications. Specifically, we consider (i) 200-atom NVT QMD simulation for liquid $\rm Al_{88} Si_{12}$ alloy at $973$ K, with LDA and $\Gamma$-point for Brillouin zone integration; (ii) structural atomic relaxation for a 52-atom system modeling a NH$_3$ adsorbate on a (110) TiO$_2$ surface with GGA and $4 \times 4 $ grid for Brillouin zone integration; (iii) structural cell relaxation for a   102-atom MoS$_2$ nanotube of diameter $3$ nm with GGA and $10$ points for Brillouin zone integration; (iv) single-point calculation of a 55-atom icosahedral Co nanoparticle with GGA and spin polarization; and (v) single-point calculations for a $74$-system test suite containing isolated systems such as clusters and molecules as well as extended systems such as crystals, surfaces, and nanowires, ranging from $2$ to $204$ atoms, encompassing $48$ different chemical elements and spin-polarized as well as unpolarized calculations.

\begin{figure}[h!]
\centering
\includegraphics[keepaspectratio=true,width=0.98\textwidth]{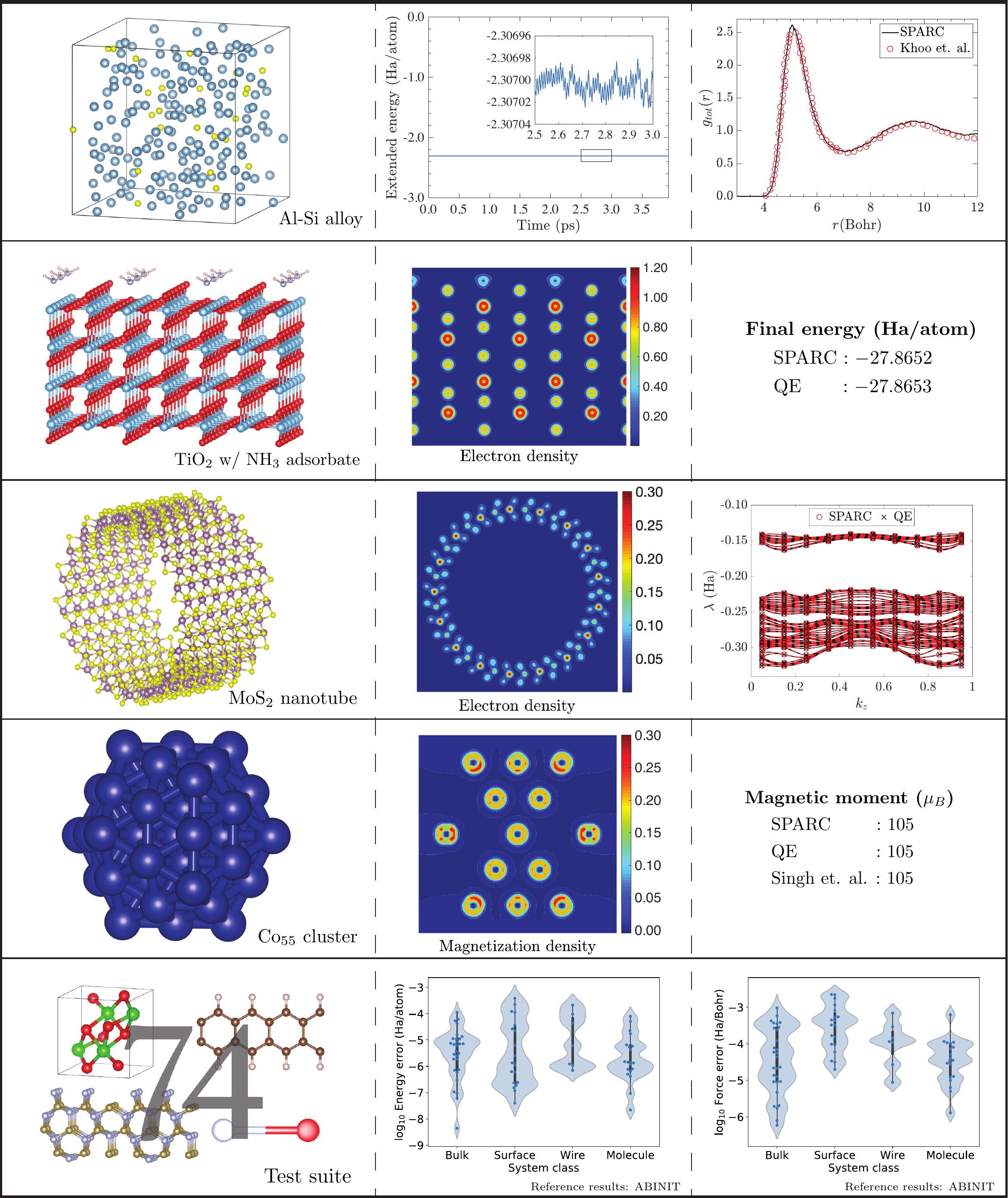}
\caption{Examples demonstrating the major functionalities of SPARC.}
\label{Fig:Examples}
\end{figure}

We employ the Perdew-Zunger parametrization for LDA \cite{PhysRevB.23.5048}, PBE variant for GGA \cite{perdew1996generalized}, and ONCV pseudopotentials \cite{hamann2013optimized,schlipf2015optimization}.  We choose mesh-sizes of $0.55$, $0.24$, $0.20$, $0.35$, and $0.18$ Bohr for the $\rm Al_{88} Si_{12}$, TiO$_2$, MoS$_2$, Co, and test suite examples, respectively. We present the results  in Fig.~\ref{Fig:Examples} (illustrations using VESTA \cite{momma2011vesta}) and compare them to established planewave codes Quantum Espresso (QE) \cite{Espresso} and ABINIT \citep{ABINIT}, as well results from the literature \cite{PhysRevB.84.214203,PhysRevB.78.245404}. It is clear that there is excellent agreement between SPARC and established planewave codes. In particular, the average difference in energy from highly converged ABINIT calculations obtained with planewave cutoff $100$ Ha for the $74$-system test suite is $\sim 2.5 \times 10^{-5}$ Ha/atom, substantially smaller than required in typical applications. Indeed, the agreement is further increased as the discretization is refined in SPARC. Overall, these examples demonstrate the capability of SPARC to obtain highly accurate results for a broad range of system compositions, configurations, and dimensionalities.

%%%%%%%%%%%%%%%%%%%%%%%%%%%%%%%%%%%%%%%%%%%%%%%%%%%%%%%%%%%%%%%%%%%%%%%%%%%%%%%%%%%%%%%%%%%%%
%%%%%%%%%%%%%%%%%%%%%%%%%%%%%%%%%%%%%%%%%%%%%%%%%%%%%%%%%%%%%%%%%%%%%%%%%%%%%%%%%%%%%%%%%%%%%
%%%%%%%%%%%%%%%%%%%%%%%%%%%%%%%%%%%%%%%%%%%%%%%%%%%%%%%%%%%%%%%%%%%%%%%%%%%%%%%%%%%%%%%%%%%%%

\section{Impact}
\label{sec:impact}
%\textbf{This is the main section of the article and the reviewers weight the description here appropriately}

%%
%%Indicate in what way new research questions can be pursued as a result of the software (if any).
%%
%%Indicate in what way, and to what extent, the pursuit of existing research questions is improved (if so).
%%
%%Indicate in what way the software has changed the daily practice of its users (if so).
%%
%%Indicate how widespread the use of the software is within and outside the intended user group.
%%
%%Indicate in what way the software is used in commercial settings and/or how it led to the creation of spin-off companies (if so).
%

Kohn-Sham DFT simulations occupy a large fraction of high-performance computing resources around the world every day \cite{nersc2014,lanlcomp2015}, a consequence of the unique insights and robust predictions they have been shown to provide.  The majority of these calculations are performed using established planewave codes \cite{VASP,CASTEP,Espresso,ABINIT,CPMD,valiev2010nwchem}. Therefore, any new implementation that is able to consistently outperform these state-of-the-art DFT codes, thereby enabling the ab initio investigation of larger length and time scales than previously accessible, with the accuracy required, stands to have significant and immediate impact. This is particularly true of a code like SPARC that is open-source with minimal dependencies so that it can be easily installed on computers large and small around the world.

\begin{figure}[h!]
\centering
\includegraphics[keepaspectratio=true,width=0.98\textwidth]{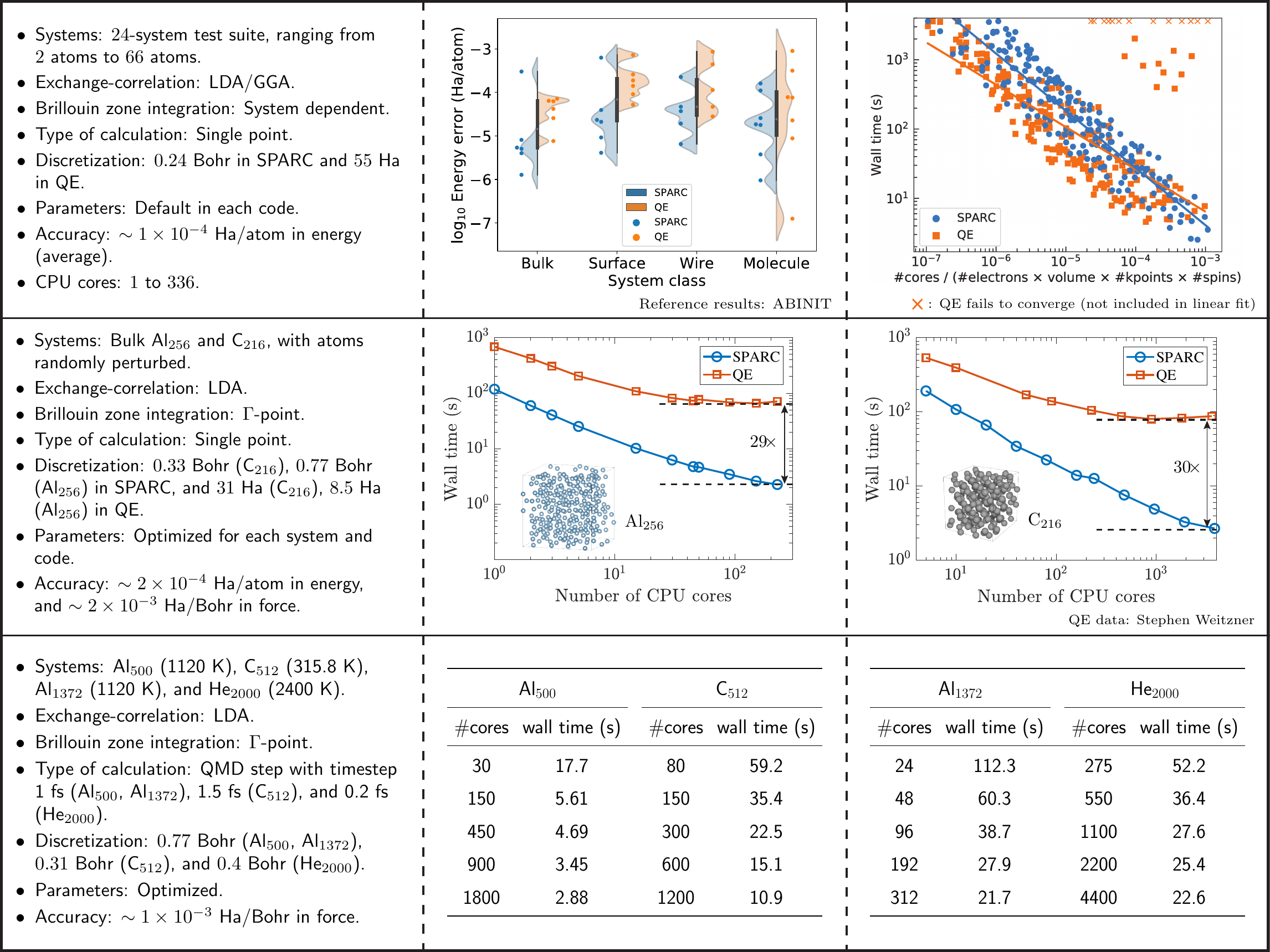}
\caption{Examples demonstrating the performance of SPARC. }
\label{Fig:AccPerform}
\end{figure}

Accordingly, we compare the accuracy and efficiency of SPARC to Quantum Espresso (QE) \cite{Espresso}, an established state-of-the-art planewave DFT code. We employ the same pseudopotentials \cite{hamann2013optimized,schlipf2015optimization} and exchange-correlation functionals \cite{PhysRevB.23.5048,perdew1996generalized} in both codes. Results and computational parameters for the study, containing a wide range of system sizes, are shown in Fig.~\ref{Fig:AccPerform}. It is clear that SPARC demonstrates comparable performance to QE on a small number of processors and increasing advantages as the number of processors grows. In particular, SPARC brings solution times down to a few seconds for systems with $\mathcal{O}(100-500)$ atoms on large-scale parallel computers, outperforming QE by more than an order of magnitude. Furthermore, SPARC achieves  QMD step times of just over $20$ seconds for the largest systems containing more than a thousand atoms, achieved on only $312$ cores for Al$_{1372}$.  For such systems and larger, the ability of SPARC to efficiently scale to many thousands of processors and more is currently limited by the subspace diagonalization step performed in each SCF iteration, which due to its cubic scaling and limited parallel scalability takes a larger fraction of wall time as the system size grows.

Going forward, we plan to first implement a structure-adapted eigensolver in SPARC to push back the cubic-scaling bottleneck, and then the DDBP method \cite{xu2018discrete} to enable strong scaling of SPARC to still larger numbers of processors, bringing down time to solution still further. The DDBP method will also enable efficient DFT calculations with hybrid functionals and the linear-scaling Spectral Quadrature (SQ) method \cite{suryanarayana2013spectral,suryanarayana2018sqdft}, which will be implemented subsequently. In order to enable the effective use of exascale computing platforms, a parallel engine for SPARC that enables highly efficient  distributed memory communication and offloading to GPUs will be completed. Moreover, machine-learning methods will be explored to improve efficiency still further. Along with these developments, we plan to implement cyclic and helical symmetry-adapted DFT formulations that allow for the highly efficient study of associated mechanical deformations as well as systems with such symmetries \cite{banerjee2016cyclic,ghosh2019symmetry,kumar2020bending}; and a coarse-grained DFT formulation that enables the study of crystal defects at realistic concentrations \citep{Phanish2012}. Indeed, many of these developments will be accelerated by using the M-SPARC code \cite{xu2020m}---same structure, algorithms, input, and output as SPARC---for rapid prototyping. 

SPARC and its variants are currently being used by multiple research groups. Moving forward, the user base is expected to grow, given the current open-source distribution, simplicity of installation and use, high accuracy, and ability to reach larger length and time scales than current state-of-the-art planewave codes. The impact thus stands to be both broad and substantial.

%%%%%%%%%%%%%%%%%%%%%%%%%%%%%%%%%%%%%%%%%%%%%%%%%%%%%%%%%%%%%%%%%%%%%%%%%%%%%%%%%%%%%%%%%%%%%
%%%%%%%%%%%%%%%%%%%%%%%%%%%%%%%%%%%%%%%%%%%%%%%%%%%%%%%%%%%%%%%%%%%%%%%%%%%%%%%%%%%%%%%%%%%%%
%%%%%%%%%%%%%%%%%%%%%%%%%%%%%%%%%%%%%%%%%%%%%%%%%%%%%%%%%%%%%%%%%%%%%%%%%%%%%%%%%%%%%%%%%%%%%
%
\section{Conclusions}
\label{sec:Conclusions}
SPARC has now become a mature code for performing real-space Kohn-Sham DFT calculations, prompting its open-source release with this publication. Currently, it can perform pseudopotential spin-polarized and unpolarized  simulations for  isolated systems such as molecules and clusters as well as extended systems such as crystals, surfaces, and nanowires, in both static and dynamic settings. SPARC is not only highly accurate, but also highly competitive with established state-of-the-art planewave codes on modest computational resources, with  increasing advantages as the number of processors increases. In particular, SPARC efficiently scales to thousands of processors, bringing solution times  for moderate-sized systems consisting of $\mathcal{O}(100-500)$ atoms to within a few seconds, making it an attractive choice for  QMD simulations in particular. Given its superior scalability, and ability to incorporate  attractive features such as linear scaling methods and variety of boundary conditions, SPARC has the potential to enable a number of new and exciting applications in science and engineering that were previously beyond reach.

%%%%%%%%%%%%%%%%%%%%%%%%%%%%%%%%%%%%%%%%%%%%%%%%%%%%%%%%%%%%%%%%%%%%%%%%%%%%%%%%%%%%%%%%%%%%
%%%%%%%%%%%%%%%%%%%%%%%%%%%%%%%%%%%%%%%%%%%%%%%%%%%%%%%%%%%%%%%%%%%%%%%%%%%%%%%%%%%%%%%%%%%%
%%%%%%%%%%%%%%%%%%%%%%%%%%%%%%%%%%%%%%%%%%%%%%%%%%%%%%%%%%%%%%%%%%%%%%%%%%%%%%%%%%%%%%%%%%%%

\section*{Conflict of Interest}
We wish to confirm that there are no known conflicts of interest associated with this publication and there has been no significant financial support for this work that could have influenced its outcome.

%%%%%%%%%%%%%%%%%%%%%%%%%%%%%%%%%%%%%%%%%%%%%%%%%%%%%%%%%%%%%%%%%%%%%%%%%%%%%%%%%%%%%%%%%%%%
%%%%%%%%%%%%%%%%%%%%%%%%%%%%%%%%%%%%%%%%%%%%%%%%%%%%%%%%%%%%%%%%%%%%%%%%%%%%%%%%%%%%%%%%%%%%
%%%%%%%%%%%%%%%%%%%%%%%%%%%%%%%%%%%%%%%%%%%%%%%%%%%%%%%%%%%%%%%%%%%%%%%%%%%%%%%%%%%%%%%%%%%%

\section*{Acknowledgements}
This work was supported by grant DE-SC0019410 funded by the U.S. Department of Energy, Office of Science. 
The work was performed in part under the auspices of the U.S. Department of Energy by Lawrence Livermore National Laboratory under Contract DE-AC52-07NA27344. Support from the Advanced Simulation \& Computing / Physics \& Engineering Models program at LLNL is gratefully acknowledged. This research was supported in part through research cyberinfrastructure resources and services provided by PACE at GT, including the Hive cluster (NSF-1828187). Time on the Quartz supercomputer was provided by the Computing Grand Challenge program at LLNL. We thank Donald Hamann for use of and assistance with a development version of the ONCVPSP code.

%%%%%%%%%%%%%%%%%%%%%%%%%%%%%%%%%%%%%%%%%%%%%%%%%%%%%%%%%%%%%%%%%%%%%%%%%%%%%%%%%%%%%%%%%%%%
%%%%%%%%%%%%%%%%%%%%%%%%%%%%%%%%%%%%%%%%%%%%%%%%%%%%%%%%%%%%%%%%%%%%%%%%%%%%%%%%%%%%%%%%%%%%
%%%%%%%%%%%%%%%%%%%%%%%%%%%%%%%%%%%%%%%%%%%%%%%%%%%%%%%%%%%%%%%%%%%%%%%%%%%%%%%%%%%%%%%%%%%%

\bibliographystyle{elsarticle-num} 
%\bibliography{SPARC}

\clearpage

\end{document}